# Anomalous Hall Effect of Reentrant Spin Glass System $Fe_{1-x}Al_x$ ($x \sim 0.3$)

Taketomo Kageyama, Natsuki Aito, Satoshi Iikubo and Masatoshi Sato[*]

*Department of Physics, Division of Material Science, Nagoya University, Furo-cho, Chikusa-ku, Nagoya 464-8602*

**Abstract**

The anomalous Hall coefficient $R_s$ has been studied for the reentrant spin glass system $Fe_{0.7}Al_{0.3}$ by measuring the magnetization $M$ and the Hall resistivity $\rho_H$. We have found that $R_s$ exhibits anomalous temperature dependence at the spin-glass transition temperature $T_G$, indicating that an additional term appears along with the beginning of the spin glass phase. The result is discussed in relation to the recent proposal of the chirality mechanism of the Hall effect in the spin glass phase.

**Keywords**

reentrant spin glass system, anomalous Hall effect, uniform chirality

[*] Corresponding author. Department of Physics, Division of Material Science, Nagoya University, Furo-cho, Chikusa-ku, Nagoya 464-8602, Japan. Tel.: +81-52-789-3537; fax: +81-52-2856.
E-mail address: msato@b-lab.phys.nagoya-u.ac.jp (M. Sato).

Observation of quite unusual behavior of the Hall resistivity $\rho_H$ of the pyrochlore molybdate $Nd_2Mo_2O_7$ has stimulated arguments on the mechanism of the anomalous Hall effect.[1,2] For ordinary ferromagnets, $\rho_H$ can be approximately described, if it is measured with the magnetic field applied perpendicularly to very thin plates of samples, by the equation $\rho_H = R_0 H + 4\pi R_s M$, where $R_0$ and $R_s$ are the ordinary and anomalous Hall coefficients, respectively, $H$ is the external magnetic field and $M$ is the magnetization.[3,4] For $Nd_2Mo_2O_7$, however, to describe the very unusual temperature($T$)- and $H$-dependences of $\rho_H$, we have to adopt an equation $\rho_H = R_0 H + 4\pi R_s M_{Mo} + 4\pi R_s' M_{Nd}$, with the Mo- and Nd- magnetizations, $M_{Mo}$ and $M_{Nd}$, respectively, though the $T$-dependence of $R_s'$ found for the system cannot simply be explained by the scattering mechanism of the Mo-4d conduction electrons by localized magnetic moments of Nd atoms.[1,5]

The data of $\rho_H$ have also been discussed in relation to the theoretical proposal that the ordering of the spin chirality $\chi$ locally defined as $\chi = \boldsymbol{S}_1 \cdot (\boldsymbol{S}_2 \times \boldsymbol{S}_3)$ for three spins $\boldsymbol{S}_1$, $\boldsymbol{S}_2$ and $\boldsymbol{S}_3$ plays an important role in determining the behavior of the anomalous Hall effect,[2,6] because the ordering of the chirality really exists in that compound below the Curie temperature. The present authors have pointed out, based on results of detailed studies of the magnetic structure determined by the neutron scattering,[7] impurity effects,[8] and other kind of data, that this chiral order does not, at least, present the main mechanism of the anomalous behavior of $\rho_H$ of $Nd_2Mo_2O_7$. However, because it is very interesting to find systems which have chirality-induced anomalous Hall resistivity, they have been searching other ferromagnets with non-trivial spin structures.[9] Spin glass systems or reentrant spin glass systems which undergo the transition to the spin glass state from the ferromagnetic one with decreasing temperature, are considered to be other candidate systems, because by the coupling between the magnetization and the spin chirality, which is considered to exist in the spin-frozen state, uniform chirality $\chi_0$ is expected to be induced, as was discussed by Kawamura's group.[10]

The reentrant spin glass system $Fe_{1-x}Al_x$ ($x \sim 0.3$) has been studied in the present work, because the transition temperature $T_G$ from the ferromagnetic state to the spin-glass one is relatively high. For this system, Shull et al.[11] presented the phase diagram obtained by the magnetization data, where we can find the reentrant spin glass phase in the region $0.27 < x < 0.31$. Motoya et al.[12] studied the change of the magnetic structure and the dynamical properties of the spin system by neutron scattering to clarify the detailed magnetic properties of the system. In the present paper, results of $\rho_H$-measurements carried out for a polycrystalline sample of $Fe_{0.7}Al_{0.3}$ are given and the relationship between the chirality $\chi$ and the Hall resistivity $\rho_H$ in the spin-glass state is discussed.

A polycrystalline rod was prepared by melting a mixture of powdered Fe and Al with the Al concentration of 0.3 sealed in a quartz capsule, in a image furnace and the rod was annealed at 700 °C for a day. Samples used in the measurements were cut from the rod. The electrical resistivity $\rho$ was measured by a conventional four-probe method from 5 K to 350 K. In four-probe measurements of the Hall resistivity $\rho_H$, thin sample plates were first cooled in zero magnetic field(ZFC) from a temperature higher than $T_G$. Then, under a fixed magnetic field $H$ applied perpendicularly to the plate, $\rho_H$ was measured by rotating the sample to reverse the field direction with respect to the sample plate to remove the background transverse voltage. With increasing $T$ stepwise up to $T$ much higher than $T_G$ and measuring $\rho_H$ at each $T$, the $T$-dependence of $\rho_H$ was taken (ZFC data), and then with decreasing $T$ stepwise, data in the field-cooled condition (FC data) were taken. To collect data at various magnetic fields, above measuring processes were repeated. The temperature dependence of the magnetization $M$ was also measured by using a SQUID magnetometer for the sample used for the $\rho_H$-measurements, at various fixed $H$ applied perpendicular to the sample plate, with increasing $T$ stepwise from 5 K to 750 K after zero field cooling (ZFC) and from 750 K to 5 K with decreasing $T$ stepwise (FC). Five-probe measurements of the Hall resistivity[13] were also carried out at various fixed temperatures, after cooling samples from the temperature $T > T_G$ in zero magnetic field (ZFC), with increasing $H$ (perpendicular to the sample plate) stepwise. The magnetization was also measured by using a SQUID magnetometer with the same conditions as used for the five-probe measurements of $\rho_H$.

Figure 1 shows the $T$-dependence of the electrical resistivity $\rho$ taken for a sample #1. The magnetizations $M$ measured for the sample #2 under the magnetic field of 100 G after zero field cooling(ZFC) and field cooling (FC) are also shown. The value of $\rho$ is small (~ 250 μΩcm) and not to be $T$-sensitive. Magnetoresistance has been found to be negligibly small. In the $M$-$T$ curve, there can be found the spin-glass like transition at around 70 K and the broad peak at around 300 K. Above 300 K, $M$ seems to obey the Curie-Weiss law. In the present paper, we use the nominal chemical formula $Fe_{0.7}Al_{0.3}$, though the Al concentration corresponds to 0.304 from the $T$-dependence of $M$ in the phase diagram of Shull et al.[11] We define the spin-glass transition temperature $T_G$ as the temperature where the difference between the magnetizations for ZFC and FC appears with decreasing $T$.

The temperature dependence of $M/H$ taken for a sample #3 under various magnetic fields after zero field cooling is shown in Fig. 2, where $M$-$T$ curves begin to decrease at around $T_G$ with decreasing $T$. Figure 3 shows the $T$ dependence of the Hall coefficients

$R_H \equiv \rho_H/H$ measured for the sample #3 by the five-probe method under various fixed magnetic fields. Each $R_H$ begins to decrease at around $T_G$ as $T$ decreases, too. In Fig. 4, the data are replotted in the form of $\rho_H/(4\pi M)$-$T$. Because the ordinary component of $\rho_H$ has been found to be rather small as compared with the anomalous one, $\rho_H/(4\pi M)$ can be considered, if the ordinary formula $\rho_H = R_0 H + 4\pi R_s M$ hold, to be approximately proportional equal to $R_s$. Then, the significant decrease of $\rho_H/(4\pi M)$ with decreasing $T$ below ~ $T_G$, suggests that an additional contribution (with negative sign) to the anomalous part appears with decreasing $T$ at around $T_G$. (We note here that $T_G$ decreases with increasing $H$.)

We have also carried out the $\rho_H$-measurements by the ordinary four-probe method by rotating samples as already stated above. In Fig. 5(a), $\rho_H$-data taken for a sample #4 under the field $H$ = 200 G by the four-probe method are compared with the $\rho_H$-data taken for the sample #3 by the five-probe method. Both data are found to be very similar. It is also noted that no appreciable difference has been observed between the data taken with the ZFC and FC conditions by the four-probe method.

In order to plot the data taken by the four-probe method in the form of $\rho_H/(4\pi M)$-$T$, we have to determine the values of $M$ which are used in the calculation of $\rho_H/(4\pi M)$. As shown in Fig. 5(b), no difference has been observed after ZFC between the values of $M$ measured before and after rotating samples by 180°. In contrast to this result, after FC significant difference between the $M$ values measured before and after rotation exists as shown in Fig. 5(b). Then, what values of $M$ are to be used in the analysis of the FC data? To answer the question, we assume that the moment system can be divided into two parts in the spin-glass state, where magnetic moments of one part can be easily aligned by the magnetic field, and those of another part is pinned to the crystal and does not rotate even when the external field is applied. Because the anomalous Hall voltage is determined as a half of the difference between the two observed before and after rotating the sample, the magnetization contributing to the Hall resistivity is just the one which can be aligned by the field. The magnetization of this part is easily obtained in both cases of ZFC and FC as the average of the values before and after the rotation. The fact that the averaged values of $M$ obtained in the FC condition are very similar to those observed in the ZFC condition indicates that the above assumption is appropriate. Then, we use these average values in the calculation of $\rho_H/(4\pi M)$. The $\rho_H/(4\pi M)$-$T$ plots thus obtained are shown in figure 6, together with the result obtained by the five-probe method in the field $H$ = 200 G. The $\rho_H/(4\pi M)$ value which is approximately equal to $R_s$ does not seem to depend on the two measuring methods used in the present study, and results shown in Figs. 4 and 6 clearly indicate that the beginning of the decrease of

$\rho_H/(4\pi M)$ or $R_s$ at around $T_G$ with decreasing $T$ is intrinsic behavior of the present system.

Now, we have found that $R_s$ or $\rho_H/(4\pi M)$ of the present system exhibits anomalous $T$-dependence at around $T_G$. What is the origin of this anomaly? As stated above, the conventional theory[3,4] predicts $\rho_H = R_0 H + 4\pi R_s M$, where $R_s$ usually depends on the reisitivity $\rho$. Based on the band model, Karplus and Luttinger,[14] for example, predicts $R_s \propto \rho^2$. Because the resistivity does not have an appreciable anomaly at $T_G$, it is difficult to explain the anomalous behavior of $\rho_H/4\pi M$ or $R_s$ as the results of the resistivity anomaly. According to the model developed by Kondo,[15] where conduction electron scattering by localized spins is considered, the magnitude of the anomalous Hall resitivity is related to the strength of the magnetic moment fluctuation. Although we do not know how the transition to the reentrant spin glass state from the ferromagnetic state affects on the fluctuation, it seems not to be easy to expect that the anomaly of $\rho_H/(4\pi M)$ is due to the decrease of the fluctuation with decreasing $T$ through $T_G$, because the fluctuation is already very small above $T_G$ but well below the Curie temperature $T_C$.

Instead, it is interesting to connect the observed anomaly with the proposal recently made by Kawamura's group[10] that the Hall resistivity is induced by the chirality mechanism in the spin-glass systems: In the spin-glass state, the chirality is frozen and by the coupling of the chirality with the ferromagnetic moment, the uniform chirality is induced in the reentrant spin-glass phase. Then, due to the existence of this uniform chirality $\chi_0$, an excess term proportional to $\chi_0$ is added to $\rho_H$ in the reentrant spin-glass phase, which gives a natural explanation of the anomalous decrease of $\rho_H/(4\pi M)$ beginning at $T_G$, with decreasing $T$.

Before summing up the results of the present studies, it may be worth noting following things. First, because we adjust the value of the transverse voltage at $H = 0$ to zero in the five probe measurements, the spontaneous Hall voltage at $H = 0$ is automatically set to be zero, even if the non-vanishing uniform chirality $\chi_0$ at $H = 0$ induces the non-zero voltage, that is, only the uniform chirality which appears as the result of the coupling with the magnetization induced by the applied field $H$ contributes to $\rho_H/(4\pi M)$. It is also true in the four-probe measurements, too. Second, Motoya et al. reported, based on the results of their neutron scattering studies, that the spin-glass state of the present system should be considered to be a cluster-glass.[12] In other word, ferromagnetic domains of spin clusters gradually develop with decreasing $T$ in the ferromagnetic state, and then, they cut the ferromagnetic network at $T_G$. If spins within a cluster ideally had the collinear structure, the chirality would not exist even below $T_G$, and then $\chi_0$ might not be induced. We think, however, that even if the cluster-glass picture is approximately appropriate to understand various observed properties, the freezing of the chirality is quite plausible.

In summary, we have studied the anomalous Hall resistivity $\rho_H$ of $Fe_{0.7}Al_{0.3}$, where anomalous *T*-dependence of $\rho_H$ has been found. Arguments on its possible connection with the chiral mechanism of the anomalous Hall resistivity have been presented.

**Acknowledgements**

The authors thank Prof. Kawamura of Osaka University for stimulating discussion on the anomalous Hall resistivity by the chirality mechanism.

**Figure captions**

Fig. 1　Temperature dependence of the electrical resistivity ρ (solid line) and the magnetizations $M$ taken with the ZFC (closed circle) and the FC (open circle) conditions under $H = 100$ G are shown for $Fe_{0.7}Al_{0.3}$.

Fig. 2　Temperature dependence of the magnetizations $M$ divided by the magnetic fields H taken with the ZFC condition.

Fig. 3　Temperature dependence of $\rho_H/H$ ($\equiv R_H$) taken by the five-probe method with various external fields $H$ is shown for $Fe_{0.7}Al_{0.3}$.

Fig. 4　Values of $\rho_H/(4\pi M)$ measured by the five-probe method are shown against $T$ at various magnetic fields.

Fig. 5　(a) Hall resistivities $\rho_H$ measured by the four-probe method are shown together with the data taken by the five-probe method. (b) Temperature dependences of $M$ measured by rotating the sample(#3) are shown both for FC and ZFC conditions. The magnitude $M$ taken with the ZFC condition does not change when the $H$-direction is reversed. $M$ values measured with the FC condition (open squares) significantly depend on the sample orientation with respect to the $H$-direction. The averages of the values at normal and the reversed sample angles are shown by closed squares. $H = 200$ G.

Fig. 6　The $\rho_H/(4\pi M)$ (~ $R_s$) values measured for $Fe_{0.7}Al_{0.3}$ by four- and five-probe methods are shown.

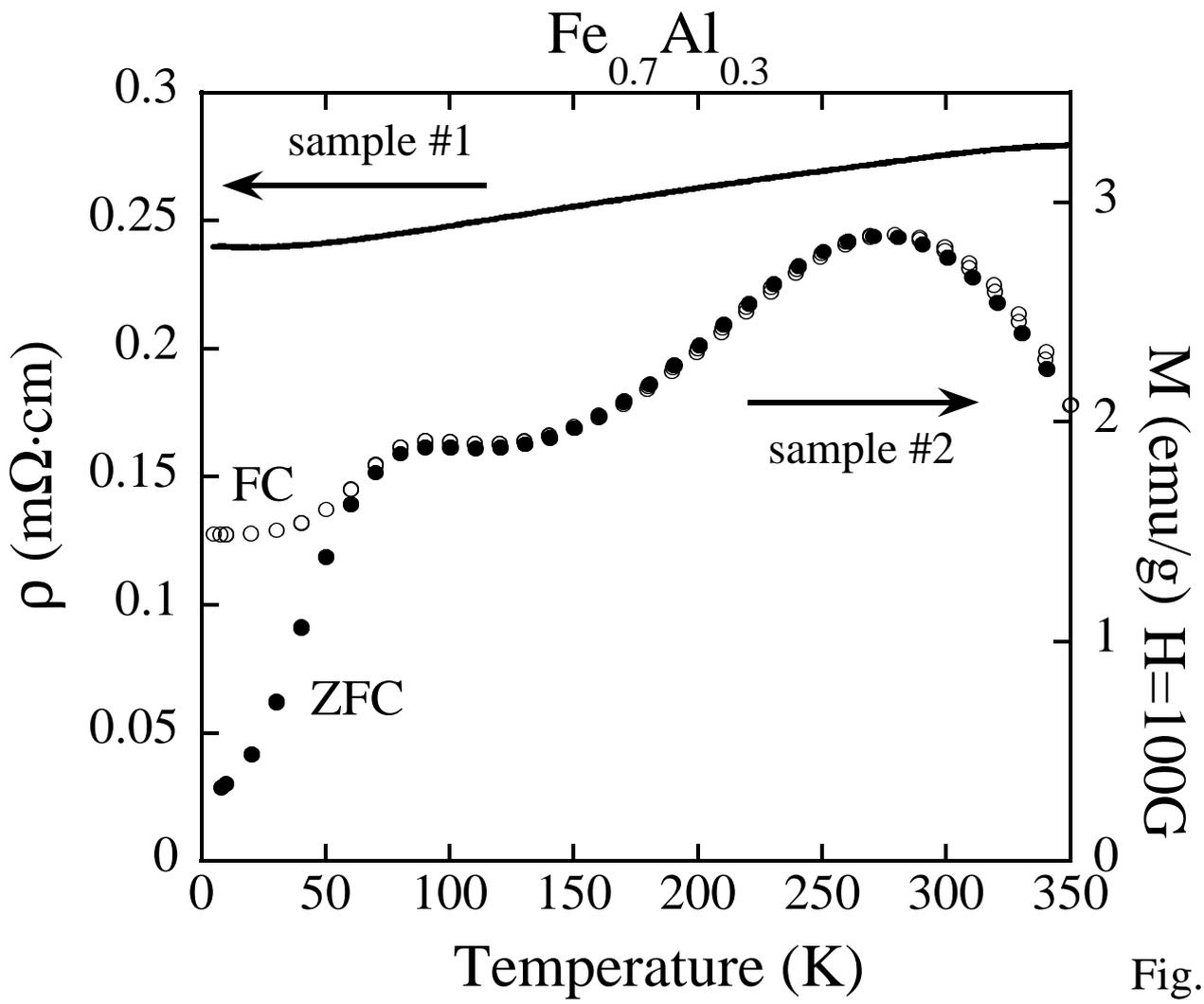

Fig. 1

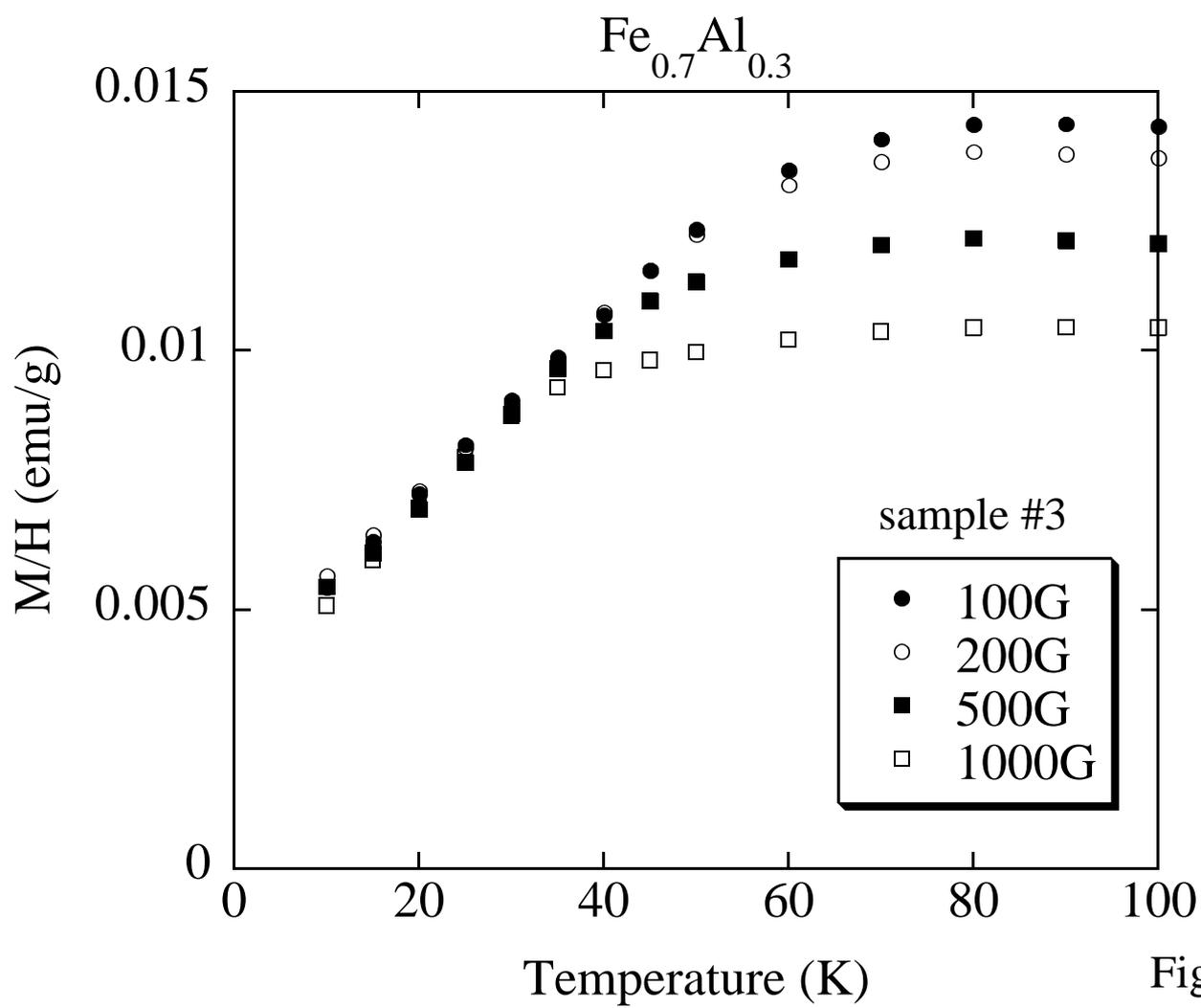

Fig. 2

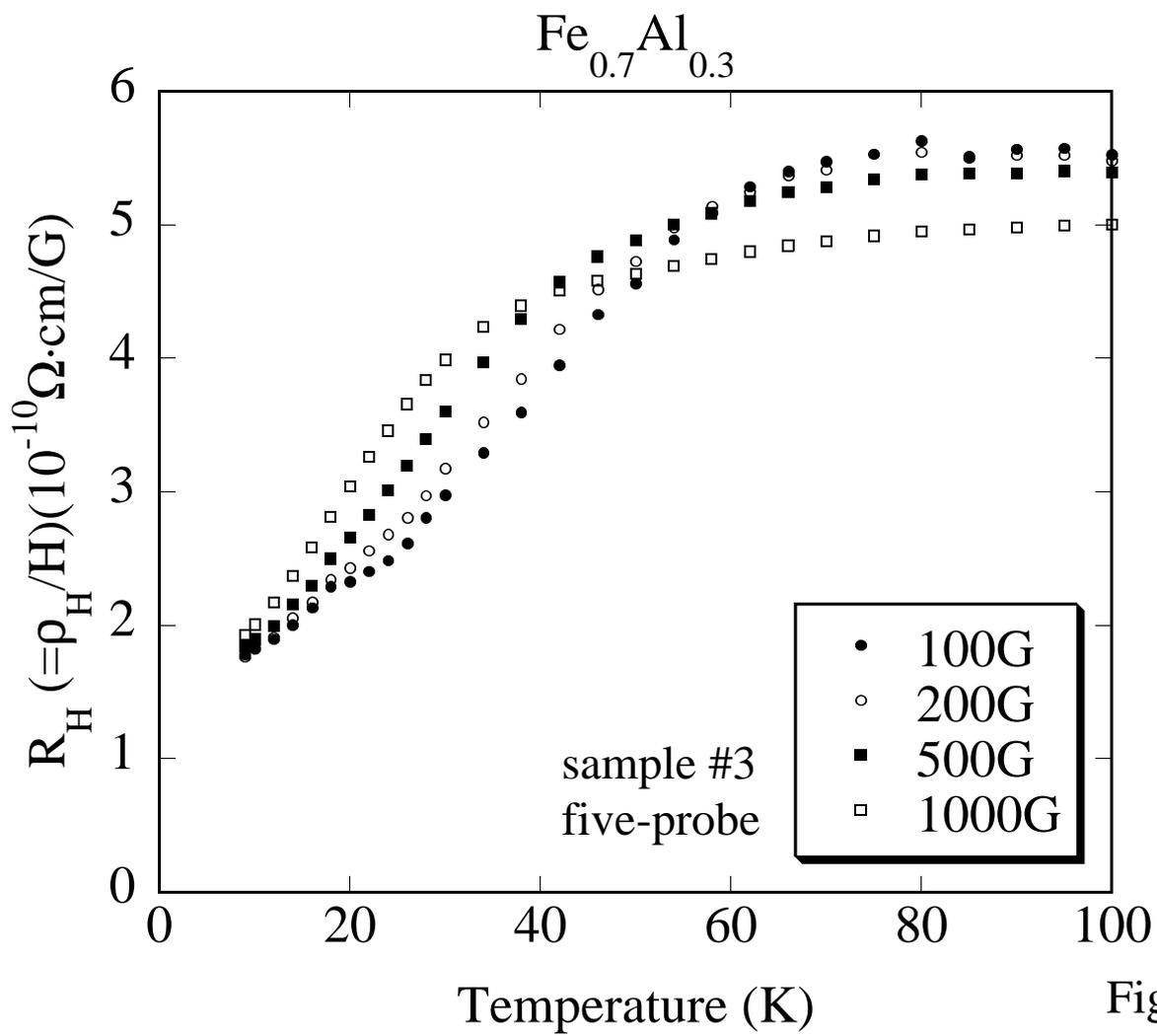

Fig. 3

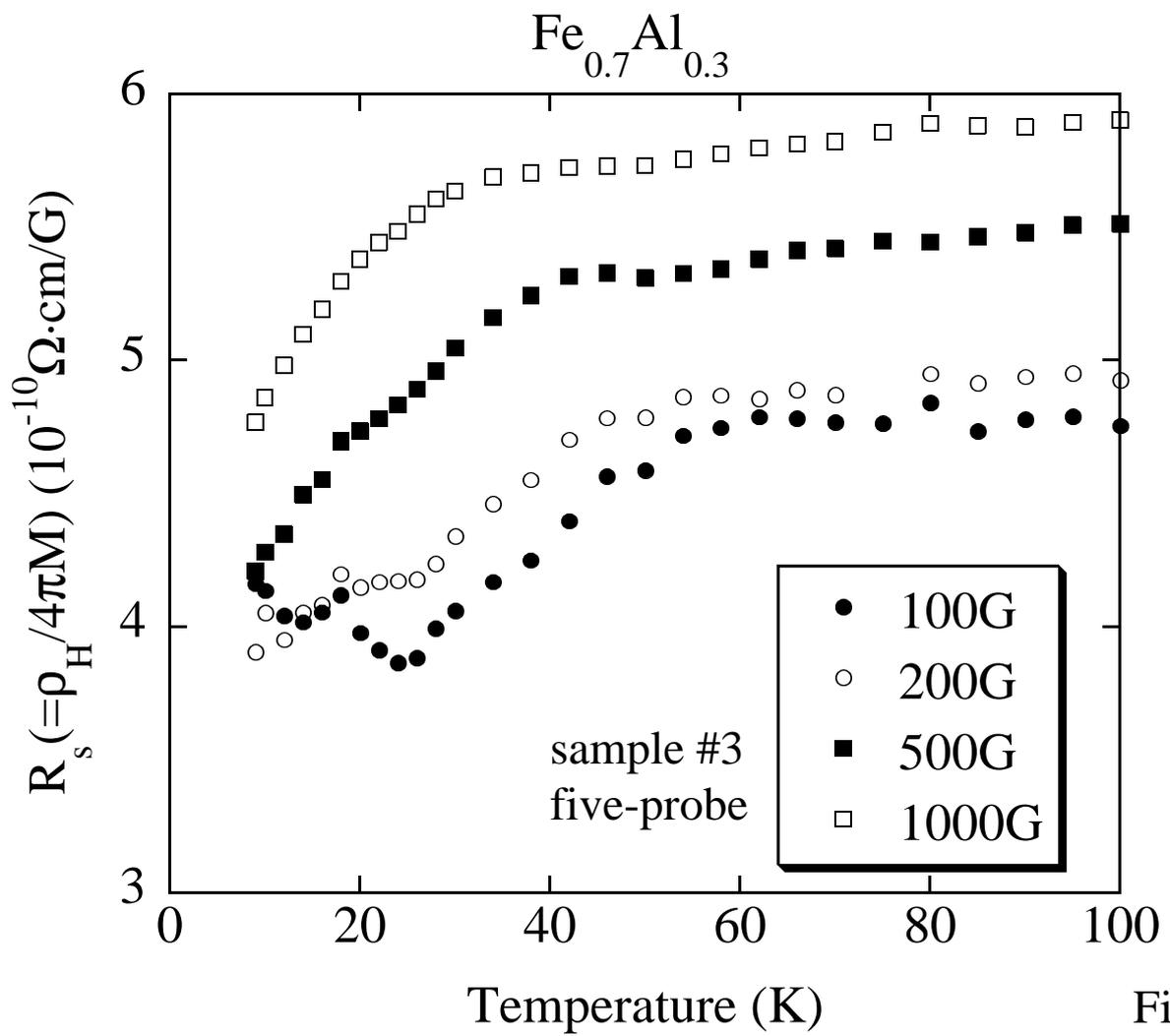

Fig. 4

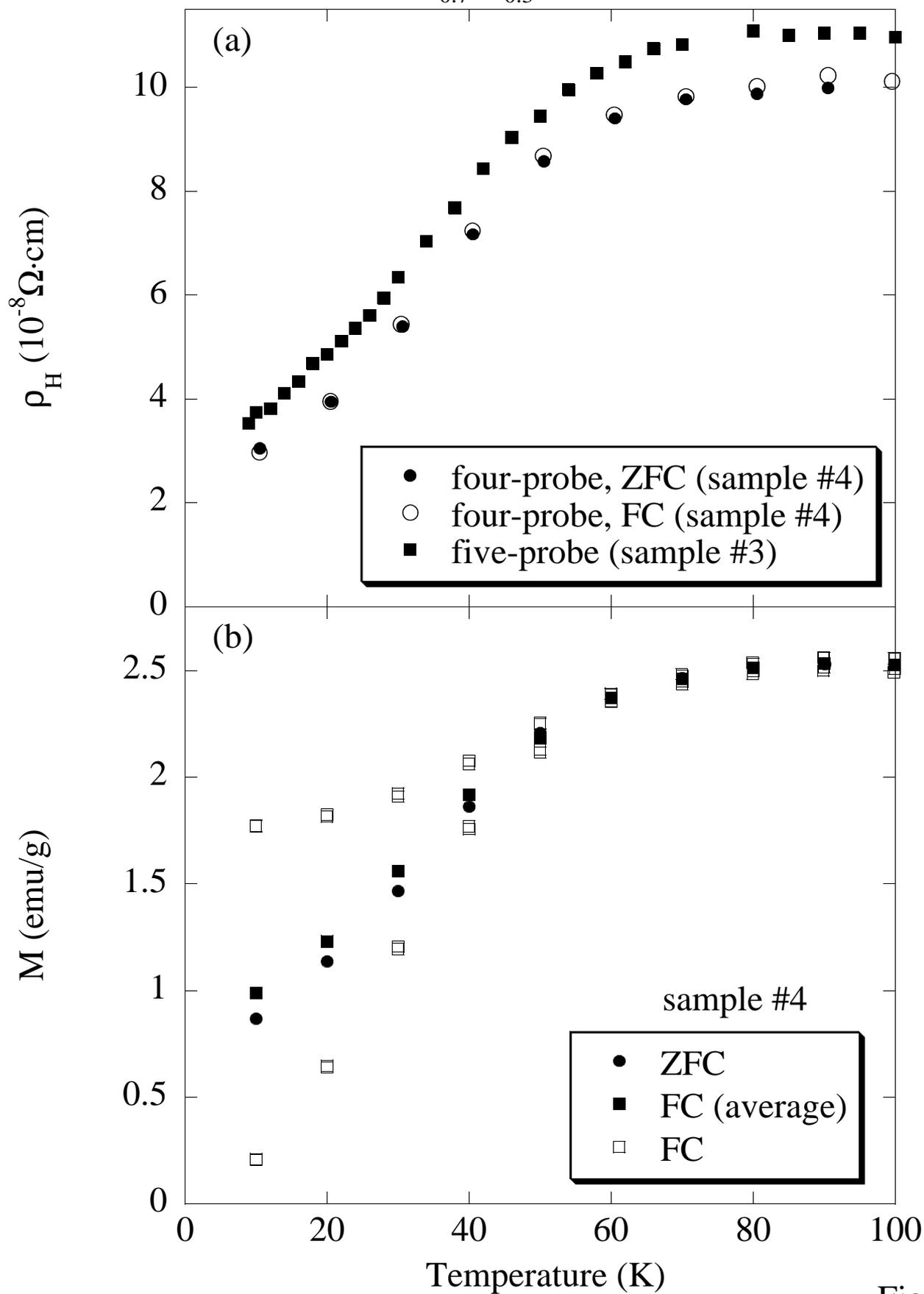

Fig. 5

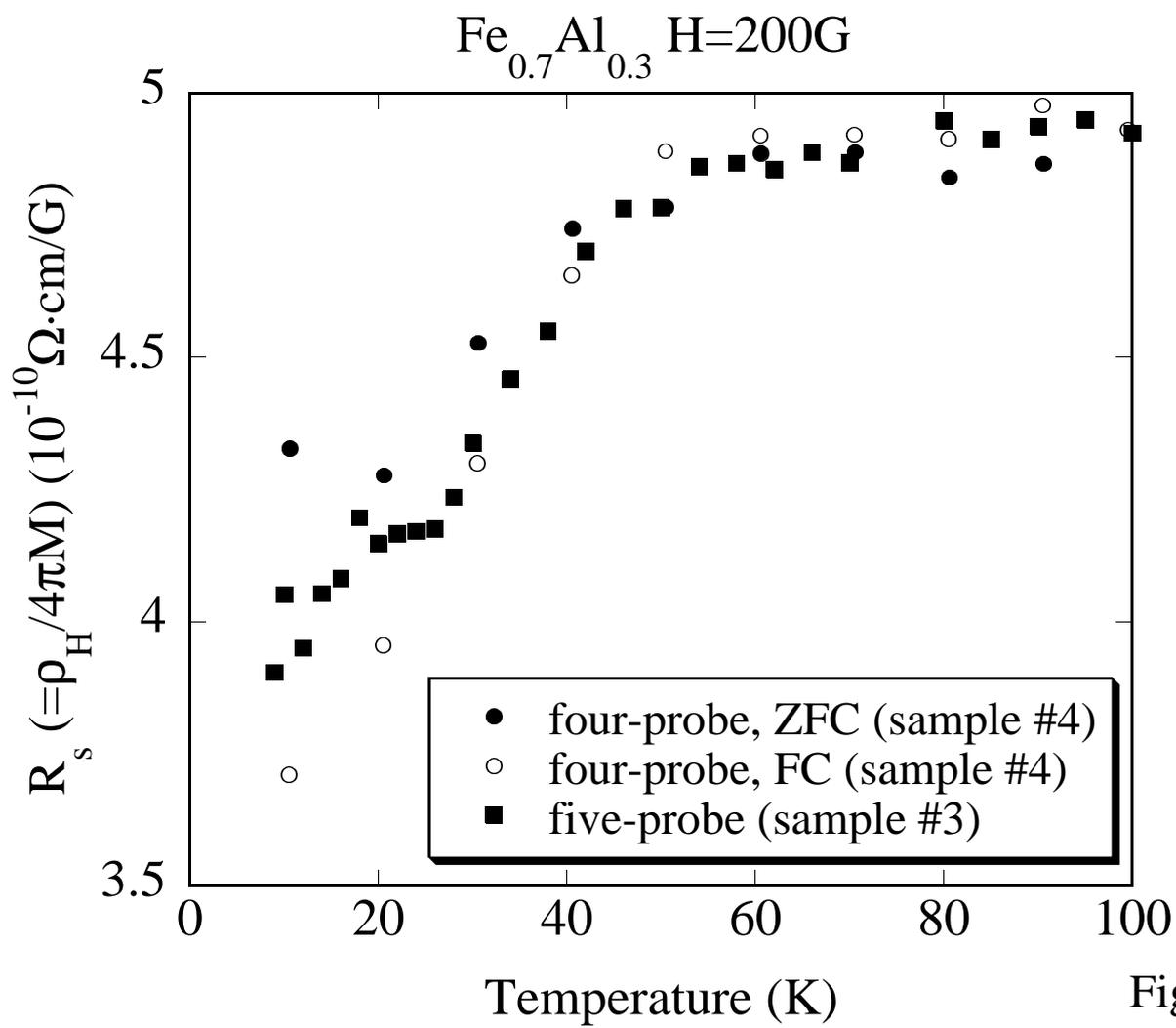

Fig. 6